\documentclass[twocolumn,amsmath,amssymb,prl,floatfix]{revtex4}
\usepackage{txfonts}
\usepackage{graphicx}
\usepackage{multirow}

\usepackage{rotating}
\usepackage{bibunits}

\newcommand{\new}[1]{#1}

\begin{document}

\begin{bibunit}[apsrev4-1]

\title{Quantum-Chemical Insights from Deep Tensor Neural Networks}
\author{Kristof T. Sch{\"u}tt$^{1}$}
\author{Farhad Arbabzadah$^1$}
\author{Stefan Chmiela$^1$}
\author{Klaus R. M{\"u}ller$^{1,2}$}
\email{klaus-robert.mueller@tu-berlin.de}
\author{Alexandre Tkatchenko$^{3,4}$}
\email{alexandre.tkatchenko@uni.lu}
\affiliation{$^1$Machine Learning Group, Technische Universit\"at Berlin, Marchstr. 23, 10587 Berlin, Germany \\
$^2$Department of Brain and Cognitive Engineering, Korea University, Anam-dong, Seongbuk-gu, Seoul 136-713, Republic of Korea \\
$^3$Fritz-Haber-Institut der Max-Planck-Gesellschaft, Faradayweg 4-6, D-14195, Berlin, Germany \\
$^4$Physics and Materials Science Research Unit, University of Luxembourg, L-1511 Luxembourg}

\begin{abstract}
Learning from data has led to paradigm shifts in a multitude of disciplines, including
web, text, and image search, speech recognition, as well as bioinformatics.
Can machine learning spur similar breakthroughs in understanding quantum many-body systems?
Here we develop an efficient deep learning approach that enables spatially and chemically
resolved insights into quantum-mechanical observables of molecular systems. 
We unify concepts from many-body Hamiltonians with purpose-designed deep tensor neural networks (DTNN),
which leads to size-extensive and uniformly accurate (1 kcal/mol) predictions in 
compositional and configurational chemical space for molecules of intermediate size. 
As an example of chemical relevance, the DTNN model reveals a classification of aromatic rings 
with respect to their stability -- a useful property that is not contained as such in the training dataset.
Further applications of DTNN for predicting atomic energies and local chemical potentials in molecules, 
reliable isomer energies, and molecules with peculiar electronic structure demonstrate the high potential of 
machine learning for revealing novel insights into complex quantum-chemical systems.
\end{abstract}

\maketitle

\begin{figure*}[h]
\includegraphics[width=0.95\textwidth]{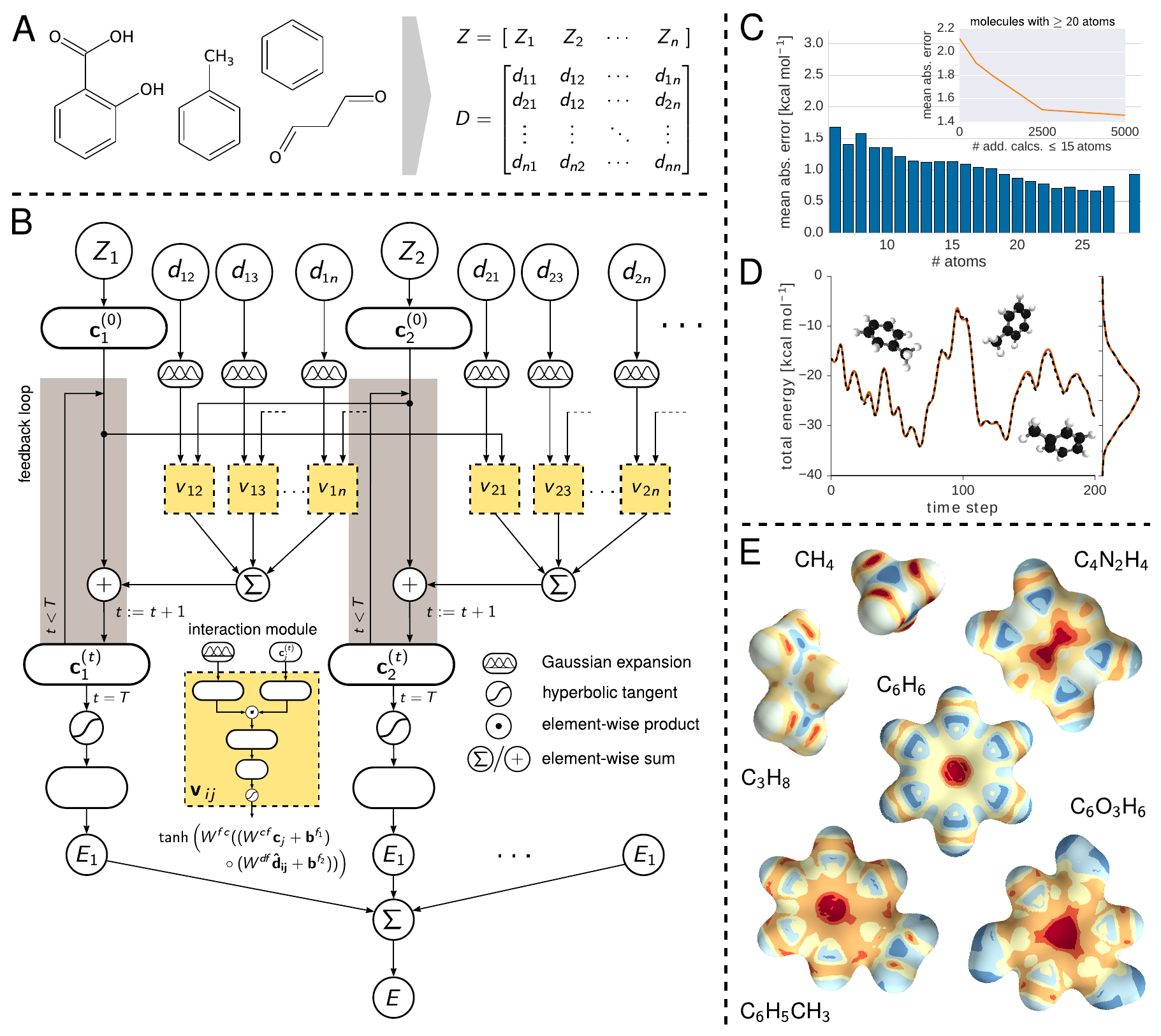}
\caption{\small \textbf{Prediction and explanation of molecular energies with a deep tensor neural network (DTNN).}
(A) Molecules are encoded as input for the neural network by a vector of nuclear charges and an inter-atomic distance matrix. This description is complete and invariant to rotation and translation.
(B) Illustration of the network architecture. Each atom type corresponds to a vector of coefficients $\mathbf{c}_i^{(0)}$ which is repeatedly refined by interactions $\mathbf{v}_{ij}$. The interactions depend on the current representation $\mathbf{c}_j^{(t)}$ as well as the distance $d_{ij}$ to an atom $j$. After $T$ iterations, an energy contribution $E_i$ is predicted for the final coefficient vector $\mathbf{c}_i^{(T)}$. The molecular energy $E$ is the sum over these atomic contributions.
(C) Mean absolute errors of predictions for the GDB-9 dataset of 129,000 molecules as a function of the number of atoms.
The employed neural network uses two interaction passes ($T=2$) and 50000 reference calculation during training. The inset shows the error of an equivalent network trained on 5000 GDB-9 molecules with 20 or more atoms, as small molecules with 15 or less atoms are added to the training set.
(D) Extract from the calculated (black) and predicted (orange) molecular dynamics trajectory of toluene. The curve on the right shows the agreement of the predicted and calculated energy distributions.
(E) Energy contribution $E_{probe}$ (or local chemical potential $\Omega_{\rm{H}}(\mathbf{r})$, see text) of a hydrogen test charge on a $\sum_i \|\mathbf{r} - \mathbf{r}_i \|^{-2}$ isosurface for various molecules from the GDB-9 dataset for a DTNN model with $T=2$.}
\end{figure*}

Chemistry permeates all aspects of our life, from the development of new drugs to the food that we consume and materials 
we use on a daily basis. 
Chemists rely on empirical observations based on creative and painstaking experimentation that leads to
eventual discoveries of molecules and materials with desired properties and mechanisms to synthesize them. 
Many discoveries in chemistry can be guided by searching large databases of experimental or computational molecular 
structures and properties by using concepts based on chemical similarity. Because the structure and properties of molecules 
are determined by the laws of quantum mechanics, ultimately chemical discovery must be based on fundamental quantum
principles. Indeed, electronic structure calculations and intelligent data analysis (machine learning, ML) 
have recently been combined aiming towards the goal of accelerated discovery of chemicals with desired 
properties~\cite{kang2009battery,norskov2009towards,hachmann2011harvard,pyzer2015high,curtarolo2013high,snyder2012finding,rupp2012fast,ramakrishnan2015big}.
However, so far the majority of these pioneering efforts have focused on the construction
of reduced models trained on large datasets of density-functional theory calculations. 
In this work, we develop an efficient deep learning approach that enables spatially and chemically resolved 
insights into quantum-mechanical properties of molecular systems beyond those trivially contained in the training dataset.
Obviously, computational models are not predictive if they lack accuracy. In addition to being interpretable, size extensive and efficient, 
our deep tensor neural network (DTNN) approach is uniformly accurate (1 kcal/mol) throughout compositional and configurational chemical space.
On the more fundamental side, the mathematical construction of the DTNN model provides statistically rigorous partitioning of 
extensive molecular properties into atomic contributions -- a long-standing challenge for quantum-mechanical calculations of molecules.

\section{Molecular Deep Tensor Neural Networks} 
It is common to use a carefully chosen representation of the problem at hand as a basis for machine learning~\cite{bishop2006pattern,ghiringhelli2015big,schutt2014represent}. For example, molecules
can be represented as Coulomb matrices~\cite{rupp2012fast,montavon2013machine,hansen2013assessment},
scattering transforms~\cite{hirn2015quantum}, 
bags of bonds (BoB)~\cite{Hansen-JCPL}, 
smooth overlap of atomic positions (SOAP)~\cite{bartok2013representing,bartok2010gaussian},
or generalized symmetry functions~\cite{behler2011atom,behler2011neural}. 
Kernel-based learning of molecular properties transforms these 
representations non-linearly by virtue of kernel functions.
In contrast, deep neural networks~\cite{lecun2015deep} are able to infer the underlying regularities and learn an efficient representation in a layer-wise fashion~\cite{montavon2011kernel}.

Molecular properties are governed by the laws of quantum mechanics, which yield the remarkable flexibility of chemical systems,
but also impose constraints on the behavior of bonding in molecules. The approach presented here utilizes the many-body Hamiltonian
concept for the construction of the DTNN architecture (see Fig.~1), embracing the principles of quantum chemistry, while maintaining the
full flexibility of a complex data-driven learning machine.
 
DTNN receives molecular structures through a vector of nuclear charges $Z$ and a matrix of atomic distances $D$ ensuring 
rotational and translational invariance by construction (Fig.~1A).
The distances are expanded in a Gaussian basis, yielding a feature vector $\hat{\mathbf{d}}_{ij}$, which 
accounts for the different nature of interactions at various distance regimes.

The total energy $E_M$ for the molecule $M$ composed of $N$ atoms is written as a sum over $N$ atomic energy contributions $E_i$, thus satisfying permutational invariance with respect to atom indexing.
Each atom $i$ is represented by a coefficient vector $\mathbf{c} \in \mathbb{R}^B$, where $B$ is the number of basis functions, or features.
Motivated by quantum-chemical atomic basis set expansions, we assign an atom type-specific descriptor vector $\mathbf{c}_{Z_i}$ 
to these coefficients $\mathbf{c}_i^{(0)}$.
Subsequently, this atomic expansion is repeatedly refined by pairwise interactions with the surrounding atoms
\begin{equation}
\mathbf{c}^{(t+1)}_i = \mathbf{c}^{(t)}_i + \sum_{j \neq i} \mathbf{v}_{ij},
\end{equation}
where the interaction term $\mathbf{v}_{ij}$ reflects the influence of atom $j$ at a distance $D_{ij}$ on atom $i$.
Note that this refinement step is seamlessly integrated into the architecture of the molecular DTNN, and is therefore
adapted throughout the learning process.
Considering a molecule as a graph, $T$ refinements of the coefficient vectors are comprised of all walks of length $T$ through the molecule ending at the corresponding atom~\cite{scarselli2009graph,duvenaud2015convolutional,suppl}.
From the point of view of many-body interatomic interactions, subsequent refinement steps $t$ correlate atomic neighborhoods with increasing complexity.

\new{
While the initial atomic representation only considers isolated atoms, the interaction terms characterize how the basis functions of two atoms overlap with each other at a certain distance. Each refinement step aims to reduce these overlaps, thereby embedding the atoms of the molecule into their chemical environment. Following this procedure, the DTNN implicitly learns an atom-centered basis that is unique and efficient with respect to the property to be predicted.}

Non-linear coupling between the atomic vector features and the interatomic distances is achieved by a 
tensor layer~\cite{socher2013recursive,sutskever2011generating}, such that the coefficient $k$ of the 
refinement is given by
\begin{equation}
v_{ijk} = \tanh \left(\mathbf{c}_{j}^{(t)} V_k \hat{\mathbf{d}}_{ij} + (W^c \mathbf{c}_{j}^{(t)})_k + (W^d \hat{\mathbf{d}}_{ij})_k  + b_k \right),
\end{equation}
where $b_k$ is the bias of feature $k$ and $W^c$ and $W^d$ are the weights of atom representation and distance, respectively.
The slice $V_k$ of the parameter tensor $V \in \mathbb{R}^{B \times B \times G}$ combines the inputs multiplicatively.
Since $V$ incorporates many parameters, using this kind of layer is both computationally expensive as well as prone to overfitting.
Therefore, we employ a low-rank tensor factorization, as described in \cite{taylor2009factored}, such that
\begin{equation}
\mathbf{v}_{ij} = \tanh \left[ W^{fc} \left( ( W^{cf} \mathbf{c}_{j} + \mathbf{b}^{f_1})  \circ (W^{df}\mathbf{\hat{d}_{ij}}  + \mathbf{b}^{f_2} ) \right) \right],
\end{equation}
where '$\circ$' represents element-wise multiplication while $W^{cf}$, $\mathbf{b}^{f_1}$, $W^{df}$, $\mathbf{b}^{f_2}$ and $W^{fc}$ 
are the weight matrices and corresponding biases of atom representations, distances and resulting factors, respectively.
As the dimensionality of $W^{cf} \mathbf{c}_{j}$ and  $W^{df} \mathbf{\hat{d}_{ij}}$ corresponds to the number of factors, 
choosing only a few drastically decreases the number of parameters, thus solving both issues of the tensor layer at once.

Arriving at the final embedding after a given number of interaction refinements, two fully-connected layers predict an energy 
contribution from each atomic coefficient vector, such that their sum corresponds to the total molecular energy $E_M$.
Therefore, the DTNN architecture scales with the number of atoms in a molecule, fully capturing the extensive nature of the energy.
All weights, biases as well as the atom type-specific descriptors were initialized randomly and trained using stochastic gradient descent~\cite{matmeth}.

\section{Learning molecular energies}
To demonstrate the versatility of the proposed DTNN, we train models with up to three interaction passes $T=3$
for both compositional and configurational degrees of freedom in molecular systems. The DTNN accuracy 
saturates at $T=3$, and leads to a strong correlation between atoms in molecules, as can be visualized 
by the complexity of the potential learned by the network (see Fig.~1E).
For training, we employ 
chemically diverse datasets of equilibrium molecular structures, as well as molecular dynamics (MD) trajectories for small 
molecules~\cite{matmeth}.
We employ two subsets of the GDB-13 database~\cite{blum2009gdb13,reymond2015chemical} referred to as GDB-7, including more than 7,000 
molecules with up to 7 heavy (C, N, O, F) atoms, and GDB-9, consisting of 129,000 molecules with up to 
9 heavy atoms~\cite{ramakrishnan2014quantum}. 
In both cases, the learning task is to predict the molecular total energy calculated with density-functional theory (DFT).
All GDB molecules are stable and synthetically accessible according to organic
chemistry rules~\cite{reymond2015chemical}. Molecular features such as functional groups or signatures include single,
double and triple bonds; (hetero-) cycles, carboxy, cyanide, amide, amine, alcohol, epoxy,
sulfide, ether, ester, chloride, aliphatic and aromatic groups. For each of the many possible
stoichiometries, many constitutional isomers are considered, each being represented only by a
low-energy conformational isomer.

As Table S2 demonstrates, DTNN achieves a mean absolute error (MAEs) of 1.0 kcal/mol on both GDB datasets,
training on 5.8k GDB-7 (80\%) and 25k (20\%) GDB-9 reference calculations, respectively~\cite{suppl}.
Fig.~1C shows the performance on GDB-9 depending on the size of the molecule.
We observe that larger molecules have lower errors because of their abundance in the training data.
\new{The per-atom DTNN energy prediction and the fact that chemical interactions have a finite
distance range means that the DTNN model will yield a constant error per atom upon increasing molecular
size. To assess the effective range of chemical interactions we have imposed a distance cutoff to interatomic interactions of $3 \AA$, yielding only a 0.1 kcal/mol increase in the error.
However, this distance cutoff restricts only the direct interactions considered in the refinement steps.
With multiple refinements, the effective cutoff increases by a factor of $T$ due to indirect interactions over multiple atoms. Given large enough molecules, so that a reasonable distance cutoff can be chosen, scaling to larger molecules will require only to have well-represented local environments.}
Along the same vein, we trained the network on a restricted subset of 5k molecules with more than 20 atoms.
By adding smaller molecules to the training set, we are able to reduce the test error from 2.1 kcal/mol to less than 
1.5 kcal/mol (see inset in Fig.~1C).
This result demonstrates that our model is able to transfer knowledge learned from small molecules to larger molecules with diverse functional groups.

While only encompassing conformations of a single molecule, reproducing MD simulation trajectories poses a radically 
different challenge to predicting energies of purely equilibrium structures.
We learned potential energies for MD trajectories of benzene, toluene, malonaldehyde and salicylic acid, carried out
at a rather high temperature of 500 K to achieve exhaustive exploration of the potential-energy surface of such small
molecules. The neural network yields mean absolute errors of 0.05 kcal/mol, 0.18 kcal/mol, 0.17 kcal/mol and 0.39 kcal/mol
for these molecules, respectively (see Table~S2). Fig.~1D shows the excellent agreement between the DFT and DTNN MD trajectory 
of toluene as well as the corresponding energy distributions. The DTNN errors are much smaller than the energy
of thermal fluctuations at room temperature ($\sim$0.6 kcal/mol), meaning that DTNN potential-energy 
surfaces can be utilized to calculate accurate molecular thermodynamic properties by virtue of Monte Carlo
simulations.

The ability of DTNN to accurately describe equilibrium structures within the GDB-9 database and 
MD trajectories of selected molecules of chemical relevance demonstrates the feasibility of developing a universal machine learning \new{architecture that can capture compositional as well as configurational degrees of freedom in the vast chemical space (see Applications section for further analysis)}. 
While the employed architecture of the DTNN is universal, the learned coefficients are different for GDB-9 and MD trajectories of single molecules.

\begin{figure}[h]
\includegraphics[width=0.48\textwidth]{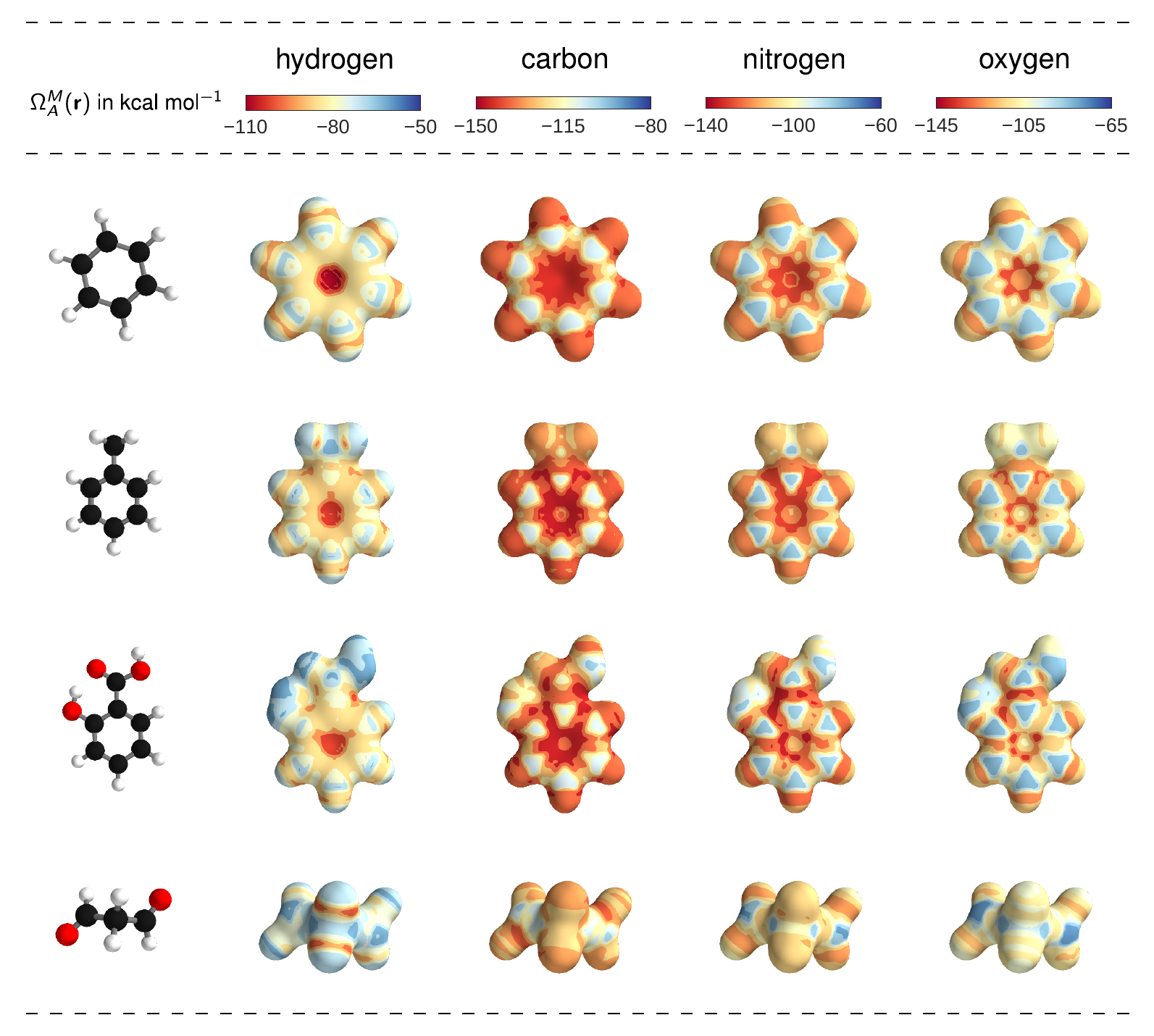}
\caption{
\small \textbf{Chemical potentials $\Omega^{M}_{A}(\mathbf{r})$ for $A=\{C,N,O,H\}$ atoms for benzene, toluene,
salicylic acid, and malondehyde.}
The isosurface was generated for $\sum_i \|\mathbf{r} - \mathbf{r}_i \|^{-2}$ = 3.8 \AA$^{-2}$ (the index $i$
is used to sum over all atoms of the corresponding molecule).}
\end{figure}

\begin{figure}[h]
\includegraphics[width=0.48\textwidth]{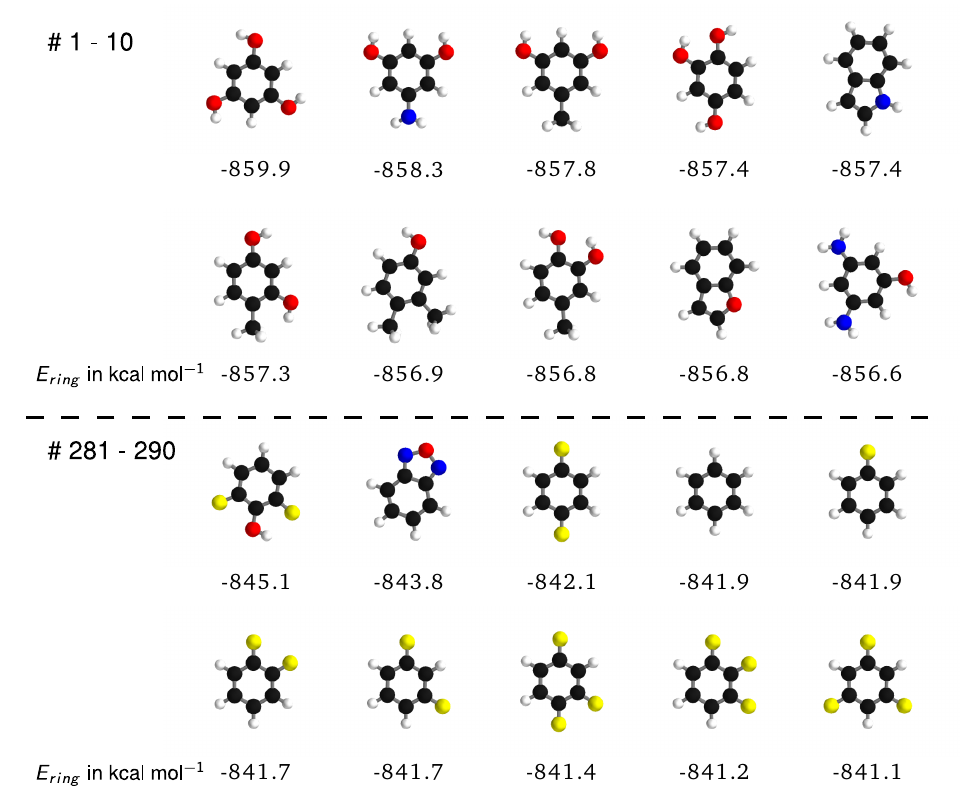}
\caption{
\small \textbf{Classification of molecular carbon ring stability.}
Shown are 20 molecules (10 most stable and 10 least stable) with respect
to the energy of the carbon ring predicted by the DTNN model.
}
\end{figure}

\begin{figure}[h]
\includegraphics[width=0.48\textwidth]{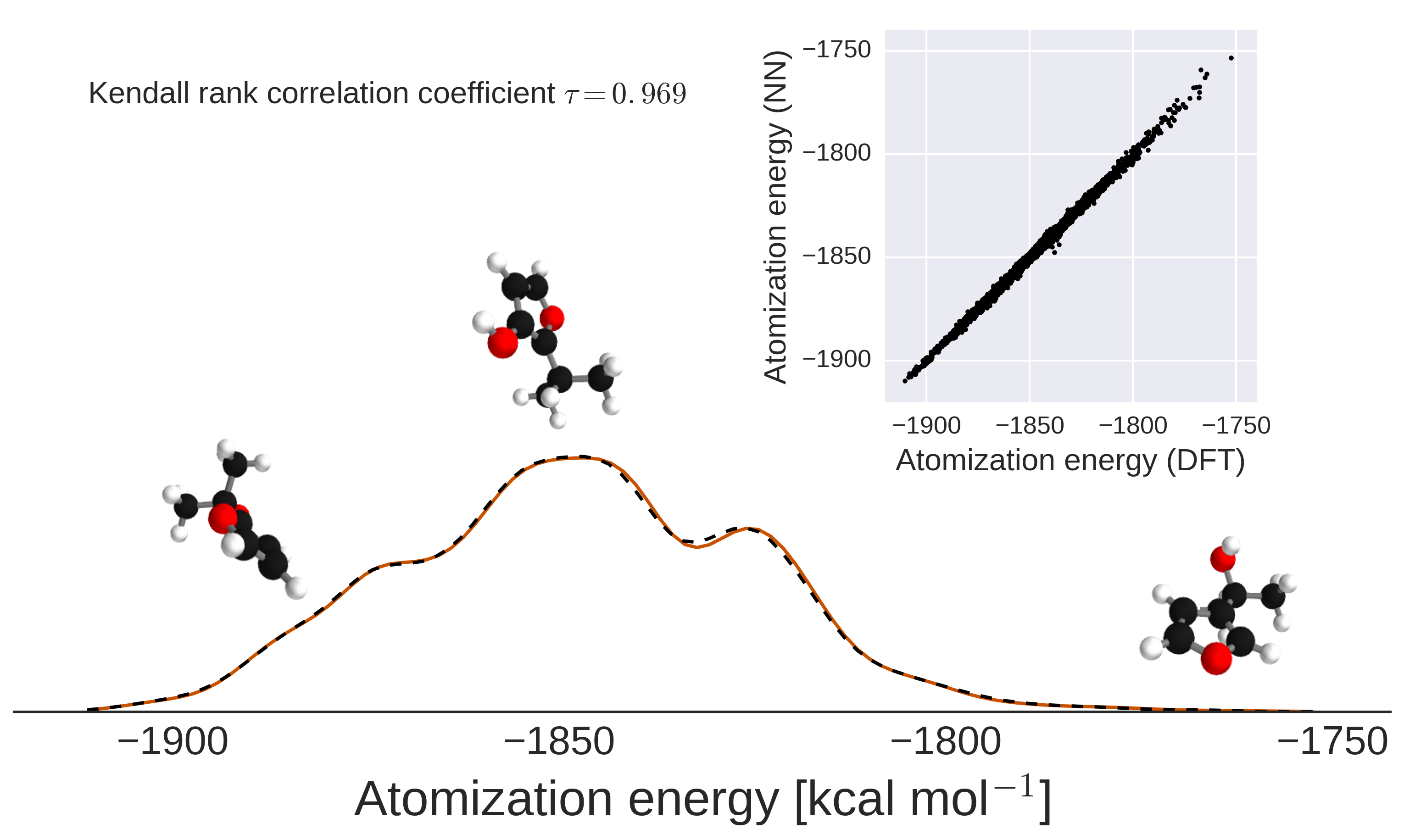}
\caption{
\small \textbf{Isomer energies with chemical formula C$_7$O$_2$H$_{10}$.}
DTNN trained on the GDB-9 database is able to acurately discriminate between
6095 different isomers of C$_7$O$_2$H$_{10}$, which exhibit a non-trivial spectrum
of relative energies.
}
\end{figure}

\section{Applications}
\subsection{Quantum-chemical insights}
Beyond predicting accurate energies, the true power of DTNN lies in its ability to provide novel quantum-chemical insights.
In the context of DTNN, we define a \textit{local chemical potential} $\Omega^{M}_{A}(\mathbf{r})$ as an energy of a 
certain atom type $A$, located at a position $\mathbf{r}$ in the molecule $M$. While the DTNN models the interatomic 
interactions, we only allow the atoms of the molecule act on the probe atom, while the probe does not influence the 
molecule~\cite{matmeth}.
The spatial and chemical sensitivity provided by our DTNN approach is shown in Fig.~1E for a variety of 
fundamental molecular building blocks. In this case, we employed hydrogen as a test charge, while the results for
$\Omega^{M}_{\rm{C,N,O}}(\mathbf{r})$ are shown in Fig.~2. Despite being trained only on total energies
of molecules, the DTNN approach clearly grasps fundamental chemical concepts such as bond saturation and different degrees of 
aromaticity. For example, the DTNN model predicts the C$_6$O$_3$H$_6$ molecule to be ``more aromatic'' than benzene
or toluene (see Fig.~1E). Remarkably, it turns out that C$_6$O$_3$H$_6$ does have higher ring stability than both benzene
and toluene and DTNN predicts it to be the molecule with the most stable aromatic carbon ring among \textit{all} 
molecules in the GDB-9 database (see Fig.~3). Further chemical effects learned by the DTNN model are shown in
Fig.~2 that demonstrates the differences in the chemical potential distribution of H, C, N, and O atoms 
in benzene, toluene, salicylic acid, and malonaldehyde. For example, the chemical potentials of different atoms over an
aromatic ring are qualitatively different for H, C, N, and O atoms -- an evident fact for a trained chemist.
However, the subtle chemical differences described by DTNN are accompanied by chemically accurate predictions
-- a challenging task for humans.

Because DTNN provides atomic energies by construction, it allows us to classify molecules by the stability of different
building blocks, for example aromatic rings or methyl groups. An example of such classification is shown
in Fig.~3, where we plot the molecules with most stable and least stable carbon aromatic rings in GDB-9.
The distribution of atomic energies is shown in Fig.~S4, while Fig.~S5 lists the full stability ranking.
The DTNN classification leads to interesting stability trends, notwithstanding the intrinsic non-uniqueness of 
atomic energy partitioning. However, unlike atomic projections employed in electronic-structure calculations, 
the DTNN approach has a firm foundation in statistical learning theory. 
In quantum-chemical calculations, every molecule would correspond to a different partitioning depending on its
self-consistent electron density. In contrast, the DTNN approach learns the partitioning on a large molecular dataset,
generating a transferable and global ``dressed atom'' representation of molecules in chemical space.
Recalling that DTNN exhibits errors 
below 1 kcal/mol, the classification shown in Fig.~3 can provide useful guidance for the chemical discovery of 
molecules with desired properties. Analytical gradients of the DTNN model with respect to chemical 
composition or $\Omega^{M}_{A}(\mathbf{r})$ could also aid in the exploration of chemical compound 
space~\cite{von2013first}. 

\subsection{Energy predictions for isomers: Towards mapping chemical space}
The quantitative accuracy achieved by DTNN and its size extensivity
paves the way to the calculation of configurational and conformational energy differences 
-- a long-standing challenge for machine learning 
approaches~\cite{rupp2012fast,montavon2013machine,hansen2013assessment,de2015comparing}.
The reliability of DTNN for isomer energy predictions
is demonstrated by the energy distribution in Fig.~4 for molecular isomers with C$_7$O$_2$H$_{10}$ chemical formula 
(a total of 6095 isomers in the GDB-9 dataset). 

\new{Training a common network for compositional and configurational degrees of freedoms requires a more complex model. Furthermore, it comes with technical challenges such as sampling and multiscale issues since the MD trajectories form clusters of small variation within the chemical compound space. As a proof of principle, we trained the DTNN to predict various MD trajectories of the C$_7$O$_2$H$_{10}$ isomers. To this end, we calculated short MD trajectories of 5000 steps each for 113 randomly picked isomers as well as consistent total energies for all equilbrium structures.
The training set is composed of the isomers in equilibrium as well as 50\% of each MD trajectory. The remaining MD calculations are used for validation and testing. Despite the vastly increased complexity, our DTNN model achieves a mean absolute error of 1.7 kcal/mol, providing a proof-of-principle demonstration of describing complex chemical spaces.}

\section{Discussion}
\new{
DTNNs provide an efficient way to represent chemical environments allowing for chemically accurate predictions.
To this end, an implicit, atom-centered basis is learned from reference \textit{ab initio} calculations.
Employing this representation, atoms can be embedded in their chemical environment within a few refinement steps.
Furthermore, DTNNs have the advantage that the embedding is built recursively from pairwise distances.
Therefore, all necessary invariances (translation, rotation, permutation) are guaranteed to be exploited by the model.}

\new{
In previous approaches, potential-energy surfaces were constructed by fitting many-body expansions with neural networks \cite{malshe2009development, manzhos2006random,manzhos2008using}.
However, these methods require a separate NN for each non-equivalent many-body term in the expansion.
Since DTNN learns a common basis in which the atom interact, higher-order interactions can obtained more efficiently without separate treament.
}

\new{Approaches like SOAP~\cite{bartok2010gaussian,bartok2013representing} or manually crafted atom-centered symmetry functions~\cite{behler2007generalized,behler2011atom,behler2011neural} are, like DTNN, based on representing chemical environments.
All these approaches have in common that size-extensivity regarding the number of atoms is achieved by predicting atomic energy contributions using a non-linear regression method (e.g., neural networks or kernel ridge regression).
However, the previous approaches have a fixed set of basis functions describing the atomic environments.
In contrast, DTNNs are able to adapt to the problem at hand in a data-driven fashion.
Beyond the obvious advantage of not having to manually select symmetry functions and carefully tune hyper-parameters of the representation, this property of the DTNN makes it possible to gain insights by analyzing the learned representation.
}

Obviously, more work is required to extend this predictive power for 
larger molecules, where the DTNN model will have to be combined with a reliable model for long-range interatomic 
(van der Waals) interactions. The intrinsic interpolation smoothness achieved by the DTNN model can also be used 
to identify molecules with peculiar electronic structure.
Fig.~S6 shows a list of molecules with the largest DTNN errors compared to reference DFT calculations. 
It is noteworthy that most molecules in this figure are characterized by unconventional bonding and the electronic 
structure of these molecules has potential multi-reference character. 
\new{The large prediction errors could stem from these molecules being not sufficiently represented by the training data.} On the other hand, DTNN predictions might turn out to be closer
to the correct answer due to its smooth interpolation in chemical space. Higher-level quantum-chemical calculations
would be required to investigate this interesting hypothesis in the future.

\section{Outlook}
We have proposed and developed a deep tensor neural network that enables understanding
of quantum-chemical many-body systems beyond properties contained in the training dataset.
The DTNN model is scalable with molecular size, efficient, and achieves uniform accuracy of
1 kcal/mol throughout compositional and configuration space for molecules of intermediate size.
The DTNN model leads to novel insights into chemical systems, a fact that we illustrated on
the example of relative aromatic ring stability, local molecular chemical potentials, relative
isomer energies, and the identification of molecules with peculiar electronic structure.

Many avenues remain for improving the DTNN model on multiple fronts. Among these 
we mention the extension of the model to increasingly larger molecules, predicting alltomic forces and
frequencies, and non-extensive electronic and optical properties.
We propose the DTNN model as a
versatile framework for understanding complex quantum-mechanical systems based on high-throughput
electronic structure calculations.

\begin{acknowledgments}
\end{acknowledgments}

\putbib[scibib]

\end{bibunit}

\begin{bibunit}[apsrev4-1]
\clearpage

\onecolumngrid
\appendix

\begin{center}
\textbf{\Large{Supplementary Materials}}
\end{center}

\section*{Materials and Methods}

\subsection*{Data}
We employ two subsets of the GDB database~\cite{blum2009gdb13}, referred to in this paper as GDB-7 and GDB-9.
GDB-7 contains 7211 molecules with up to 7 heavy atoms out of the elements C, N, O, S and Cl, saturated with hydrogen~\cite{montavon2013machine}.
Similarly, GDB-9 includes 133,885 molecules with up to 9 heavy atoms out of C, O, N, F~\cite{ramakrishnan2014quantum}.
Both data sets include calculations of atomization energies employing density functional theory~\cite{DFT} with the PBE0~\cite{PBE0} and B3LYP~\cite{B3,LYP,vosko1980accurate,stephens1994ab,becke1993phys} exchange-correlation potential, respectively.

The molecular dynamics trajectories are calculated at a temperature of 500~K and resolution of 0.5fs using density functional theory with the PBE exchange-correlation potential~\cite{PBE}.
The data sets for benzene, toluene, malonaldehyde and salicylic acid consist of 627k, 442k, 993k and 320k time steps, respectively.
In the presented experiments, we predict the potential energy of the MD geometries.

\subsection*{The deep tensor neural network model}

The molecular energies of the various data sets are predicted using a deep tensor neural network.
The core idea is to represent atoms in the molecule as vectors depending on their type and to subsequently refine the representation by embedding the atoms in their neighborhood.
This is done in a sequence of interaction passes where the atom representations influence each other in a pair-wise fashion.
While each of these refinements depends only on the pair-wise atomic distances, multiple passes enable the architecture to also take angular information into account.
Due to this decomposition of atomic interactions, an efficient representation of embedded atoms is learned following quantum chemical principles.

In the following, we describe the deep tensor neural network step-by-step, including hyper-parameters used in our experiments.

\begin{enumerate}
\item \textbf{Assign initial atomic descriptors} \\
We assign an initial coefficient vector to each atom $i$ of the molecule according to its nuclear charge $Z$:
\begin{equation}
\mathbf{c}_i^{(0)} = \mathbf{c}_{Z_i} \in R^B,
\end{equation}
where B is the number of basis functions.
All presented models use atomic descriptors with 30 coefficients.
We initialize each coefficient randomly following $\mathbf{c}_{Z} \sim \mathcal{N}(0, 1/\sqrt{B})$.

\item \textbf{Gaussian feature expansion of the inter-atomic distances} \\
The inter-atomic distances $d_{ij}$ are spread across many dimensions by a uniform grid of Gaussians
\begin{equation}
\hat{\mathbf{d}}_{ij} = \left[ \exp \left( -\frac{(d_{ij} - (\mu_{min}+ k\Delta \mu))^2 }{ 2\sigma^2} \right) \right]_{0 \leq k \leq \mu_{max}/\Delta \mu},
\end{equation}
with $\Delta \mu$ being the gap between two Gaussians of width $\sigma$.

In our experiments, we set both to $0.2 \,$\r{A}.
The center of the first Gaussian $\mu_{min}$ was set to $-1$, while $\mu_{max}$ was chosen depending on the range of distances in the data ($10 \,$\r{A} for GDB-7 and benzene, $15 \,$\r{A} for toluene, malonaldehyde and salicylic acid and $20 \,$\r{A} for GDB-9).

\item \textbf{Perform $T$ interaction passes}
Each coefficient vector $\mathbf{c}^{(t)}_i$, corresponding to atom $i$ after $t$ passes, is corrected by the interactions with the other atoms of the molecule:
\begin{equation}
\mathbf{c}^{(t+1)}_i = \mathbf{c}^{(t)}_i + \sum_{j \neq i} \mathbf{v}_{ij}.
\end{equation}
Here, we model the interaction $v$ as follows:
\begin{equation}
\mathbf{v}_{ij} = \tanh \left( W^{fc} ( ( W^{cf} \mathbf{c}_{j} + \mathbf{b}^{f_1})  \circ (W^{df}\mathbf{\hat{d}_{ij}}  + \mathbf{b}^{f_2} ) ) \right),
\end{equation}
where the circle ($\circ$) represents the element-wise matrix product.
The factor representation in the presented models employs 60 neurons.

\item \textbf{Predict energy contributions} \\
Finally, we predict the energy contributions $E_i$ from each atom $i$. Employing two fully-connected layers, for each atom a scaled energy contribution $\hat{E}_i$ is predicted:
\begin{equation}
\mathbf{o}_i = \tanh (W^{out_1} \mathbf{c}^{(T)}_i + \mathbf{b}^{out_1})
\end{equation}
\begin{equation}
\hat{E}_i = W^{out_2} \mathbf{o}_i + \mathbf{b}^{out_2}
\end{equation}
In our experiments, the hidden layer $\mathbf{o}_i$ possesses 15 neurons.
To obtain the final contributions, $\hat{E}_i$ is shifted to the mean $E_\mu$ and scaled by the standard deviation $E_{std}$ of the energy per atom estimated on the training set.
\begin{equation}
E_i = (\hat{E}_i + E_\mu) E_{std}
\end{equation}
This procedure ensures a good starting point for the training.
\item \textbf{Obtain the molecular energy} $E = \sum_i E_i$
\end{enumerate}

The bias parameters as well as $W^{out_2}$ are initially set to zero. All other weight matrices are initialized drawing from a uniform distribution according to~\cite{glorot2010understanding}.

The deep tensor neural networks have been trained for 3000 epochs minimizing the squared error, using stochastic gradient descent with 0.9 momentum and a constant learning rate~\cite{lecun2012efficient}.
The final results are taken from the models with the best validation error in early stopping.

\subsection*{Computational cost of training and prediction}

All DTNN models were trained and executed on an NVIDIA Tesla K40 GPU. 
The computational cost of the employed models depends on the number of reference calculations, the number of interaction passes as well as the number of atoms per molecule. 
The training times for all models and data sets are shown in Table \ref{tab:traintime}, ranging from 6 hours for 5.768 reference calculations of GDB-7 with one interaction pass, to 162 hours for 100.000 reference calculations of the GDB-9 data set with 3 interaction passes.

On the other hand, the prediction is instantaneous: all models predict examples from the employed data sets in less than 1 ms.
Fig. \ref{fig:predtime} shows the scaling of the prediction time with the number of atoms and interaction layers. Even for a molecule with 100 atoms, a DTNN with 3 interaction layers requires less than 5 ms for a prediction.

The prediction as well as the training steps scale linearly with the number of interaction passes and quadratically with the number of atoms, since the pairwise atomic distances are required for the interactions.
For large molecules it is reasonable to introduce a distance cutoff (future work).
In that case, the DTNN will also scale linearly with the number of atoms.

\subsection*{Computing the local potentials of the DTNN}

\begin{figure}[h!]
\centering
\includegraphics[width=0.8\textwidth]{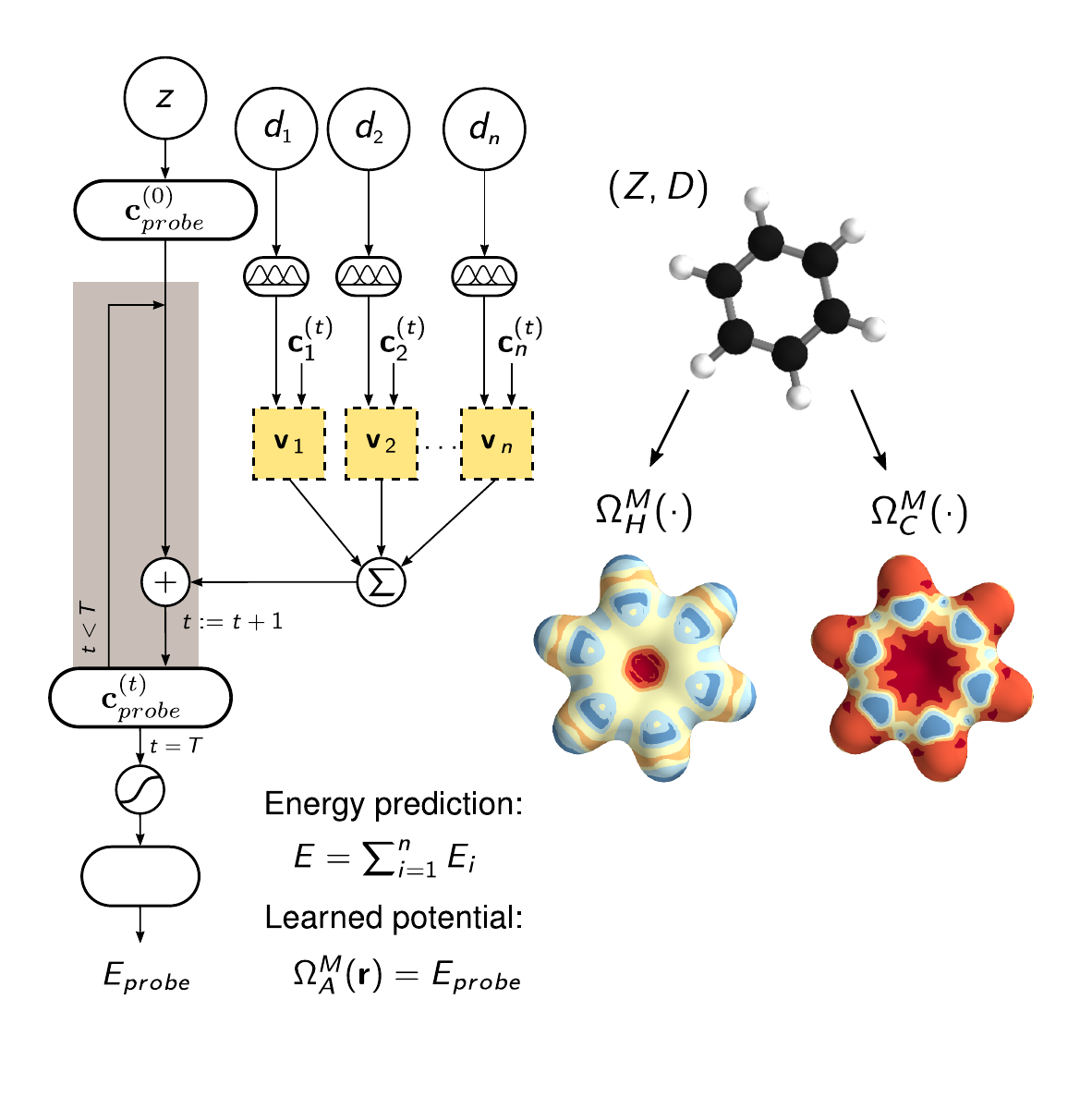}
\caption{ \textbf{Illustration of how the surface plots are obtained from a trained network as shown in Fig. 1.} The deep network can be interpreted as representing a local potential $\Omega_A^M(\mathbf{r})$ created by the atoms of the molecule. Putting a probe atom A with nuclear charge $z$ at a position $\mathbf{r}$ described by the distances to the atoms of the molecule $d_1, \dots, d_n$ yields an energy $E_{probe}$.}
\label{fig:explanationpot}
\end{figure}

Given a trained neural network as described in the previous section, one can extract the coefficients vectors $\mathbf{c}^{(t)}_i$ for each atom $i$ and each interaction pass $t$ for a molecule of interest.
From each final representation $\mathbf{c}^{(T)}_i$, the energy contribution $E_i$ of the corresponding atom to the molecular energy can be obtained.
Instead, we let the molecule act on a probe atom, described by its charge $z$ and the pairwise distances $d_{1}, \dots, d_{n}$ to the atoms of the molecule:
\begin{equation}
\mathbf{c}^{(t+1)}_{probe} = \mathbf{c}^{(t)}_{probe} + \sum_{j=1}^{n} \mathbf{v}_{j},
\end{equation}
with $\mathbf{v}_{j} = \tanh \left( W^{fc} ( ( W^{cf} \mathbf{c}_{j} + \mathbf{b}^{f_1})  \circ (W^{df}\mathbf{\hat{d}}_{j}  + \mathbf{b}^{f_2} ) ) \right)$.
While this is equivalent to how the coefficient vectors of the molecule are corrected, here, the molecule does not get to be influenced by the probe.
Now, the energy of the probe atom is predicted as usual from the final representation $\mathbf{c}^{(T)}_{probe}$.
Interpreting this as a local potential $\Omega_A^M(\mathbf{r})$ generated by the molecule, we can use the neural network to visualize the learned interactions as illustrated in Fig. \ref{fig:explanationpot}.
The presented energy surfaces show the potential for different probe atoms plotted on an isosurface of $\sum_{i=1}^n d_i^{-2}$.
We used Mayavi~\cite{ramachandran2011mayavi} for the visualization of the surfaces.

\subsection*{Computing an alchemical path with the DTNN}

The alchemical paths in Fig. \ref{fig:alchemy} were generated by gradually moving the atoms as well as interpolating between the initial coefficient vectors for changes of atom types.
Given two nuclear charges $A, B$, the coefficient vector for any charge $Z_i = \alpha_i A + (1-\alpha)B$ with $0 \leq \alpha \leq 1$ is given by
\begin{equation}
\mathbf{c}_{Z_i} = \alpha_i \mathbf{c}_{A} + (1-\alpha_i) \mathbf{c}_{B}.
\end{equation}
Similarly, in order to add or remove atoms, we introduce fading factors $\beta_1, \dots, \beta_n \in [0, 1]$ for each atom. This way, influences on other atoms
\begin{equation}
\mathbf{c}^{(t+1)}_i = \mathbf{c}^{(t)}_i + \sum_{j \neq i} \beta_j v(\mathbf{c}^{(t)}_j, D_{ij})
\end{equation}
as well as energy contributions to the molecular energy $E = \sum_{i=1}^n \beta_i E_i$ can be faded out.

\clearpage

\section*{Supplementary Text}

\subsection*{Discussion of the results}

Table \ref{tab:results} shows the mean absolute (MAE) and root mean squared errors (RMSE) as well as standard errors over five randomly drawn training sets. For GDB-9 and the MD data sets, 1k reference calculations were used as validation set for early stopping, while the remaining data was used for testing. In case of GDB-7, validation and test set contained 10\% of the reference calculations each.
With respect to the mean absolute error, chemical accuracy can be achieved on all employed data sets using models with two or three interactions passes.

\begin{figure}[h!]
\centering
\includegraphics[width=0.7\textwidth]{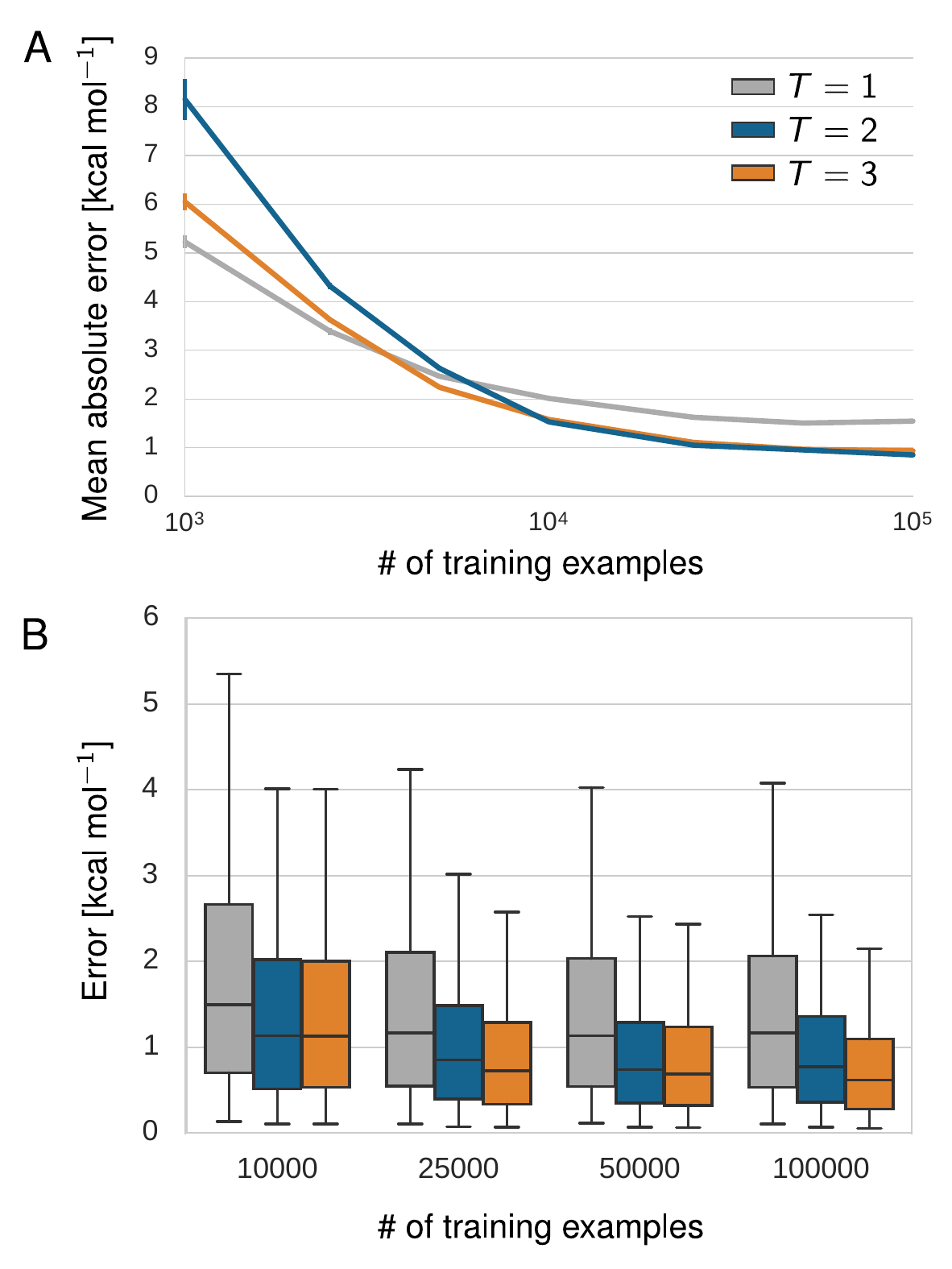}
\caption{\textbf{Chemical compound space. Errors depending on the size of the training set for models with $T=1,2,3$ interaction passes trained on GDB-9.} 
(A) Mean absolute error of neural networks depending on the number of training examples. Error bars correspond to standard errors over five repetitions. For more than 5k examples, the error bars vanish due to standard errors below 0.05 kcal/mol. (B) Error distribution for models trained on 10k, 25k, 50k and 100k training examples. The box spans between the 25\% and 75\% quantiles, while the whiskers mark the 5\% and 95\% quantiles.}\label{fig:ccslc}
\end{figure}

\begin{figure}[h!]
\centering
\includegraphics[width=0.7\textwidth]{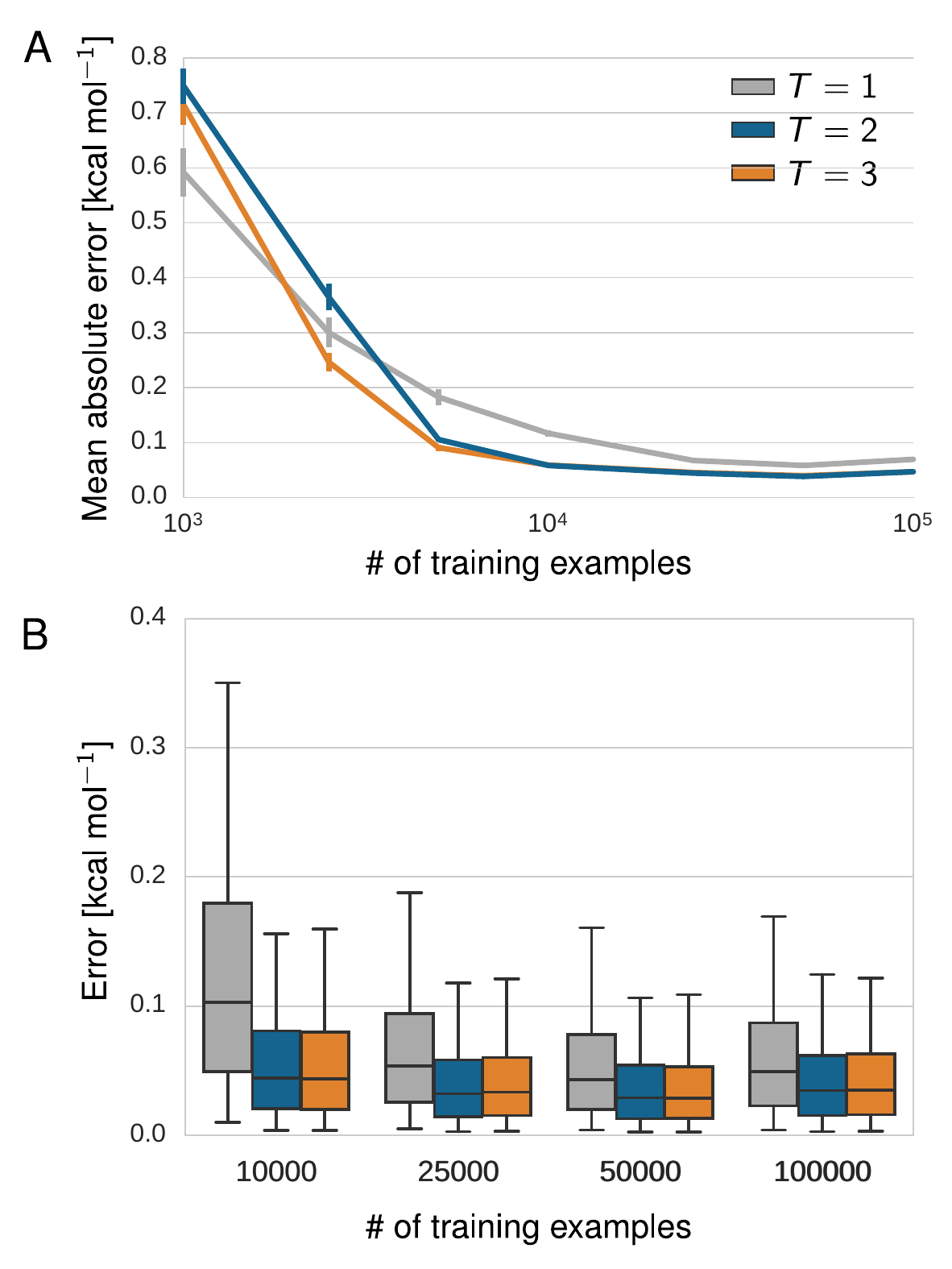}
\caption{\textbf{Molecular dynamics. Errors depending on the size of the training set for models with $T=1,2,3$ interaction passes trained on the Benzene data set.} 
(A) Mean absolute error of neural networks depending on the number of training examples. Error bars correspond to standard errors over five repetitions. For more than 10k examples, the error bars vanish due to standard errors below 0.01 kcal/mol. (B) Error distribution for models trained on 10k, 25k, 50k and 100k training examples. The box spans between the 25\% and 75\% quantiles, while the whiskers mark the 5\% and 95\% quantiles.}\label{fig:mdlc}
\end{figure}

Figs. \ref{fig:ccslc} and \ref{fig:mdlc} show the dependence of the performance on the number of training examples for the benzene MD data set and GDB-9, respectively.
In both learning curves (A), an increase from 1.000 to 10.000 training examples reduces the error drastically while another increase to 100.000 examples yields comparatively small improvement.
The error distributions (B) show that models with two and three interaction passes trained on at least  25.000 GDB-9 references calculations predict 95\% of the unknown molecules with an error of 3.0 kcal/mol or lower.
Correspondingly, the same models trained on 25.000 or more MD reference calculations of benzene predict 
95\% of the unknown benzene configurations with a maximum error lower than 1.3 kcal/mol.

Beyond a certain number of reference calculations, the models with one interaction pass perform significantly worse in all theses respects.
Thus, multiple interaction passes indeed enrich the learned feature representation as demonstrated by the increased predictability of previously unseen molecules.

\subsection*{Relations to other deep neural networks}

Deep learning has lead to major advances in computer vision, language processing, speech recognition and other applications~\cite{lecun2015deep}.
In our model, we embed the atom type in a vector space $R^B$.
This idea is inspired by word embeddings (word2vec) employed in natural language processing~\cite{mikolov2013distributed}.
In order to model inter-atomic effects, we need to represent the influence of an atom represented by $\mathbf{c}_j$ at the distance $d_{ij}$.
To account for multiple regimes of atomic distances as well as different dimensionality of the two inputs, we apply the Gaussian feature mapping described above.
Similar approaches have been applied to the entries of the Coulomb matrix for the prediction of molecular properties before~\cite{montavon2013machine}.
A natural way to connect distance and atom representation is a tensor layer as used in text generation~\cite{sutskever2011generating}, reasoning~\cite{socher2013reasoning} or sentiment analysis~\cite{socher2013recursive}.
For an efficient computation as well as regularization, we employ a factorized tensor layer, corresponding to a low-rank approximation of the tensor product~\cite{taylor2009factored}.

Convolutional neural networks have been applied to images, speech and text with great success due to their ability to capture local structure~\cite{ciresan2012multi,krizhevsky2012imagenet,lecun1995convolutional,hinton2012deep,sainath2015deep,collobert2008unified}.
In a convolution layer, local filters are applied to local environments, e.g., image patches, extracting features relevant to the classification task.
Similarly, local correlations of atoms may be exploited in a chemistry setting.
The atom interaction in our model can indeed be regarded as a non-linear generalization of a convolution.
In contrast to images however, atoms of molecules are not arranged on a grid.
Therefore, the convolution kernels need to be continuous.
We define a function $C^t: R^3 \rightarrow R^B$ yielding $\mathbf{c}_i^t = C^t(\mathbf{r}_i)$ at the atom positions.
Now, we can rewrite the interactions as
\begin{equation}
C^{t+1}(\mathbf{r}_i)_k = C^{t}(\mathbf{r}_i)_k + \sum_{j \neq i} h(f(\mathbf{r}_j)_k g(\| \mathbf{r}_j-\mathbf{r}_i \|)_k),
\end{equation}
with 
\begin{equation}
f(\mathbf{r}_j)=W^{cf} C^{t}(\mathbf{r}_j) + \mathbf{b}^{f_1},
\end{equation}
\begin{equation}
g(D_{ij}) = W^{df} \mathbf{\hat{d}_{ij}} + \mathbf{b}^{f_2},
\end{equation}
\begin{equation}
h(x) = \tanh( W^{fc} \mathbf{x} ).
\end{equation}

For $h$ being the identity, the sum is equivalent to a discrete convolution.

\clearpage

\section*{Figures}

\clearpage

\clearpage

\clearpage

\begin{figure}[h!]
\centering
\includegraphics[width=\textwidth]{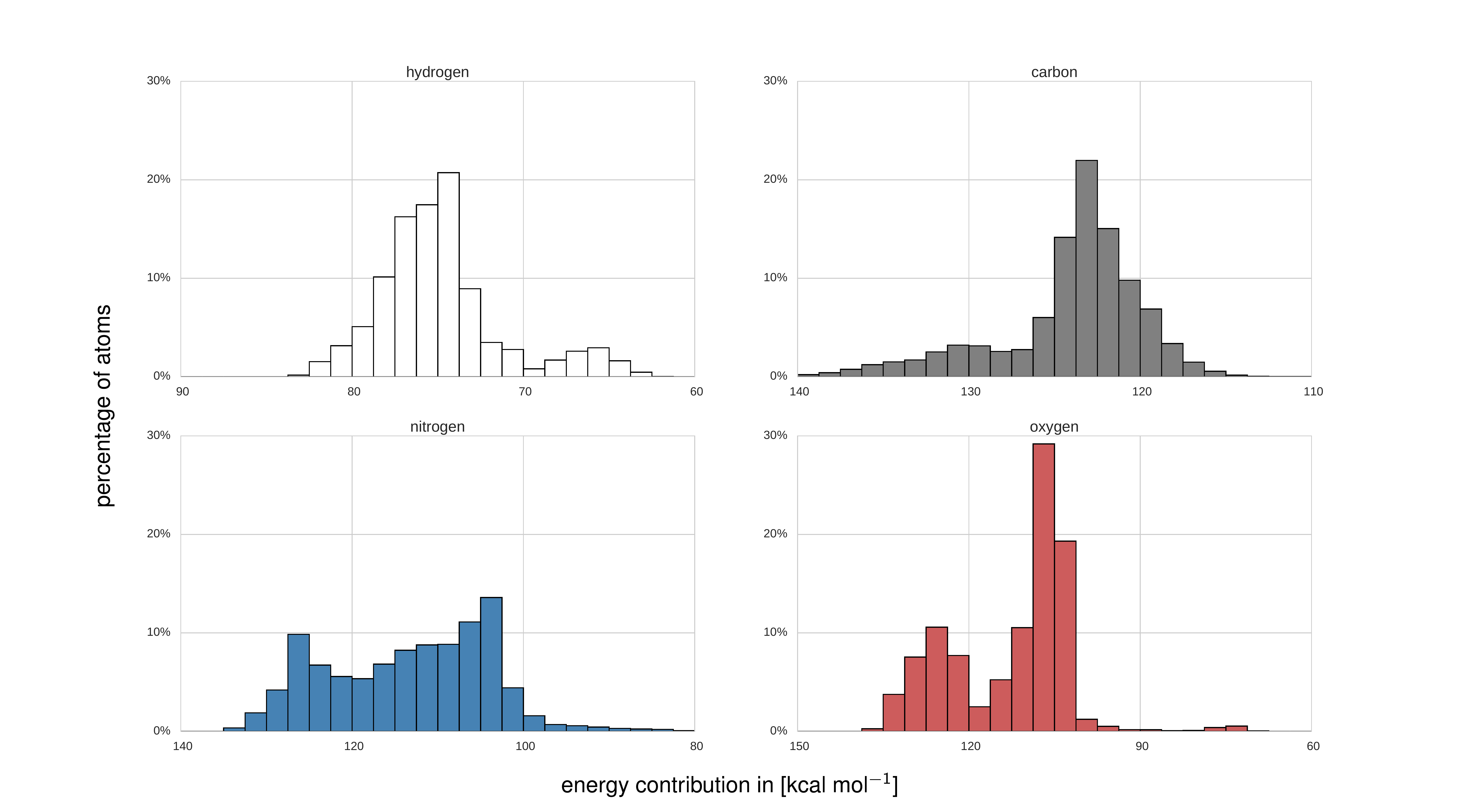}
\caption{\textbf{Distribution of atomic energy contributions $E_i$ in the GDB-9 data set.} The energy contributions were predicted using the GDB-9 model with two interaction passes trained on 50k reference calculations.}
\end{figure}
\clearpage

\begin{figure}[h!]
\centering
\includegraphics[width=\textwidth]{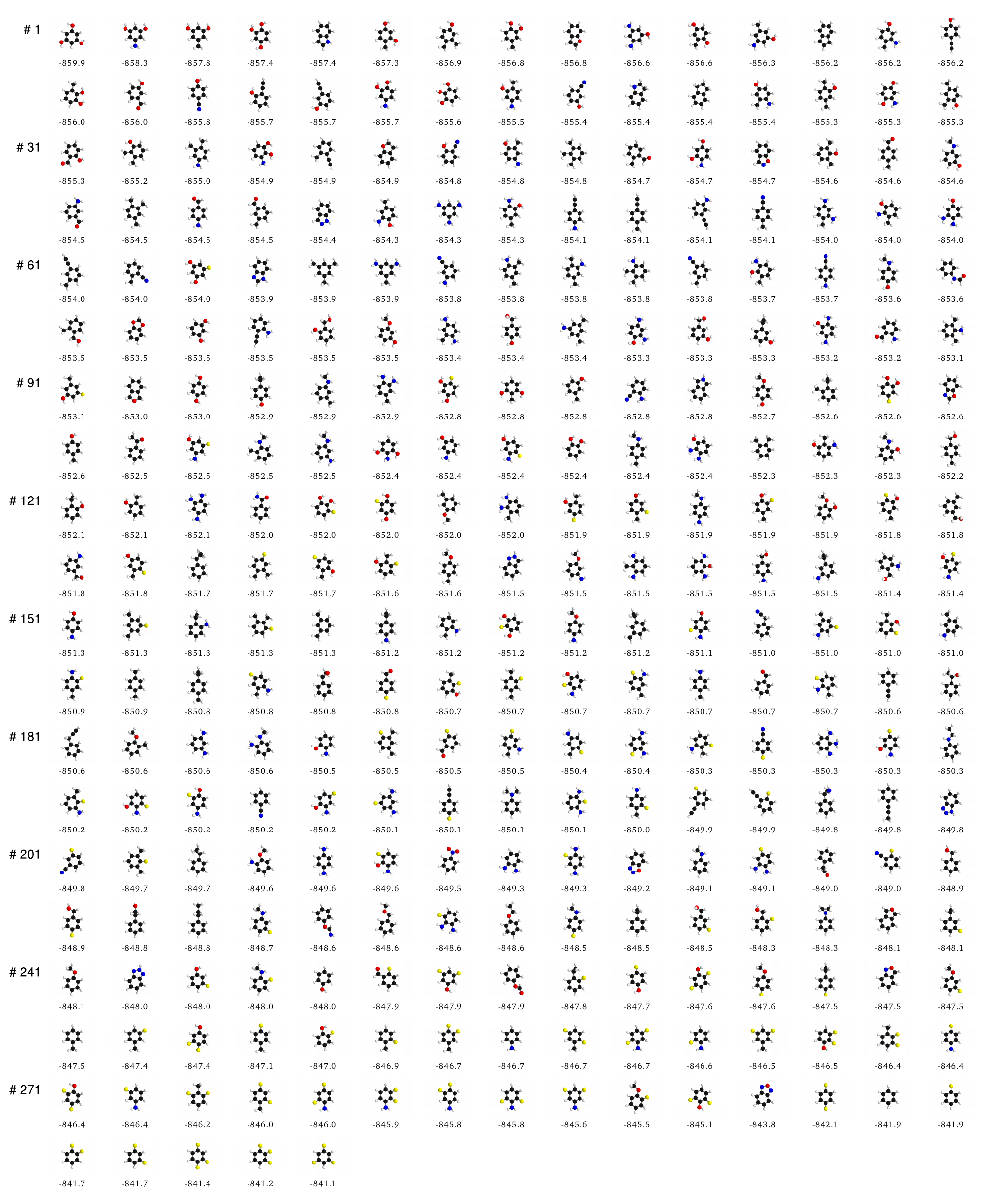}
\caption{\textbf{List of 6-membered carbon rings ordered by the sum of energy contributions of the ring atoms.} The energy contributions were predicted using the GDB-9 model with three interaction passes trained on 50k reference calculations. Energy contributions are given in kcal/mol.}
\end{figure}
\clearpage

\begin{figure}[h!]
\centering
\includegraphics[width=\textwidth]{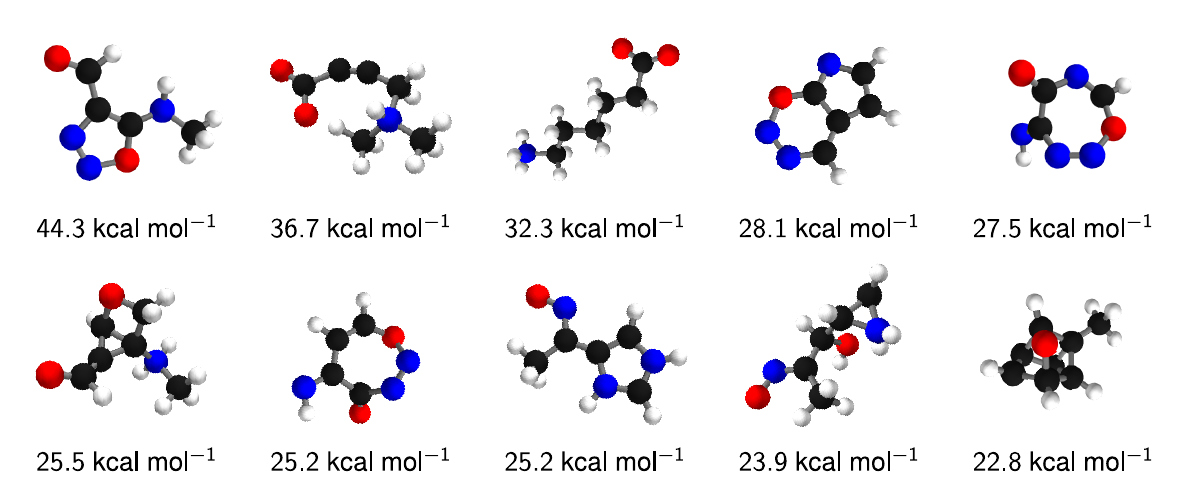}
\caption{\textbf{Top-10 largest prediction errors on the GDB-9 model with two interaction passes trained on 50k reference calculations.}}
\end{figure}
\clearpage

\begin{figure}[h!]
\centering
\includegraphics[width=0.6\textwidth]{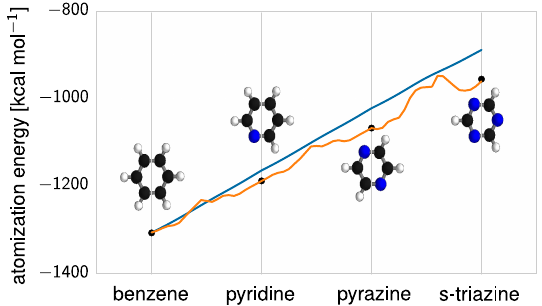}
\caption{\small  \textbf{An alchemical path of the DTNN trained on 50k GDB-9 reference calculations with $T=2$.}
The DTNN model is able to smoothly create, remove and move atoms as well as continuously change their element-specific characteristics.
A path leading from benzene to s-triazine was computed by only changing, removing and changing types of atoms (blue). In the second path (orange), atoms were also moved to the new equilibrium positions.
The black dots mark the energy of DFT reference calculations.}\label{fig:alchemy}
\end{figure}
\clearpage

\begin{figure}[h!]
\centering
\includegraphics[width=\textwidth]{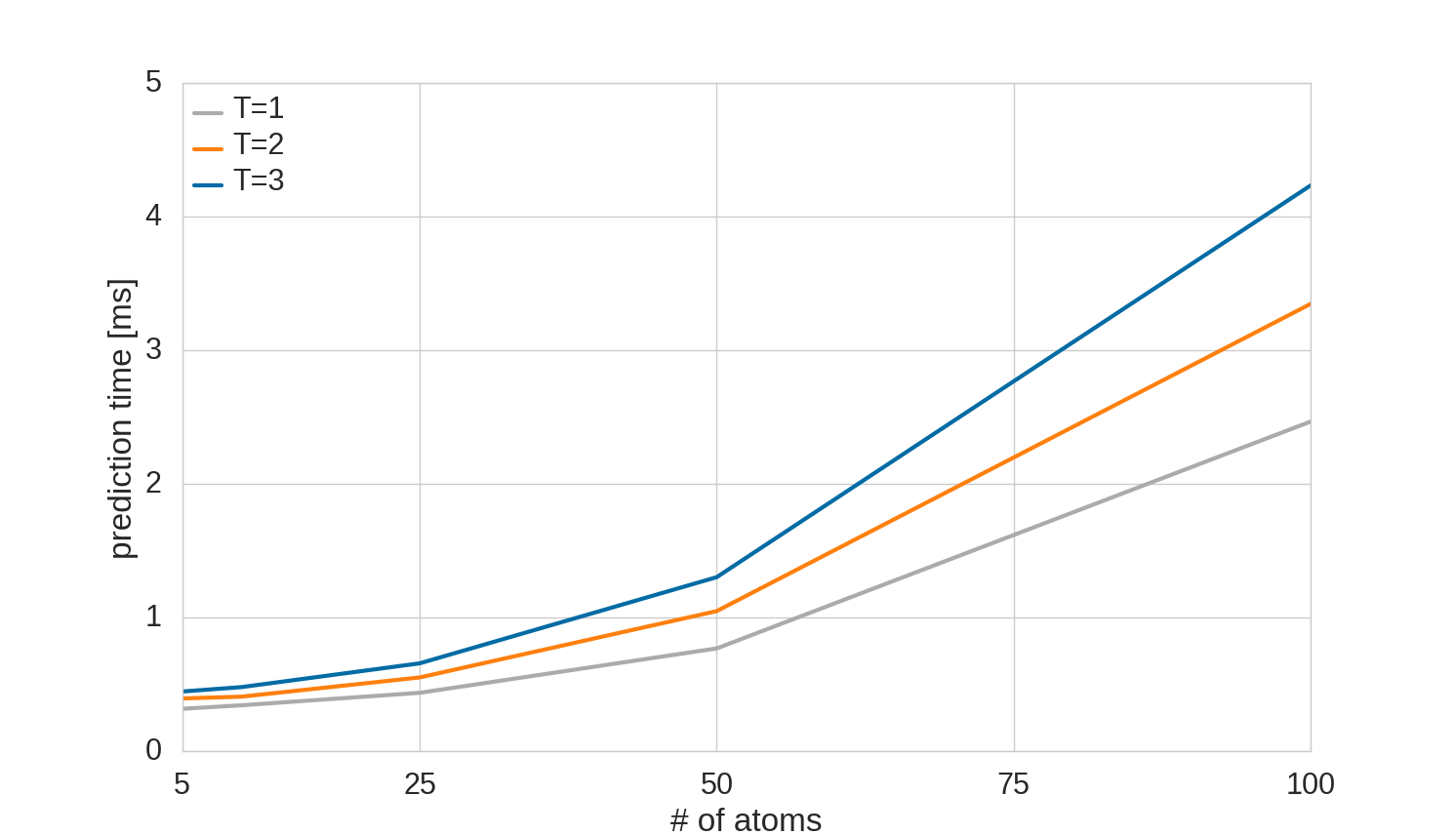}
\caption{\small  \textbf{Prediction time needed for a molecule depending on the number of atoms and number of interaction passes $T$ of the employed DTNN.} All predictions were computed on an NVIDIA Tesla K40 GPU.}\label{fig:predtime}
\end{figure}
\clearpage

\section*{Tables}

\begin{table}[h]
\centering
\begin{tabular}{lrrrr}
Data set & \# training examples & $T=1$ & $T=2$ & $T=3$ \\[0.5ex]
\hline & \\ [-1.5ex]
GDB-7 &  5768 & 6  &  7  &  8 \\[0.5ex]
\hline & \\ [-1.5ex]
GDB-9 & 25k & 28  &  35  &  42\\[0.5ex]
& 50k & 55  &  71  &  82 \\[0.5ex]
& 100k & 110  &  139  &  162\\[0.5ex]
\hline & \\ [-1.5ex]
Benzene & 25k  & 21  &  27  &  32\\[0.5ex]
& 50k & 44  &  53  &  61 \\[0.5ex]
& 100k & 84  &  104  &  121 \\[0.5ex]
\hline & \\ [-1.5ex]
Toluene & 25k & 24  &  27  &  32 \\[0.5ex]
& 50k & 45  &  55  &  64 \\[0.5ex]
& 100k & 88  &  108  &  127 \\[0.5ex]
\hline & \\ [-1.5ex]
Malonaldehyde & 25k & 21  &  25  &  29\\[0.5ex]
& 50k & 41  &  52  &  59 \\[0.5ex]
& 100k & 85  &  106  &  117\\[0.5ex]
\hline & \\ [-1.5ex]
Salicylic acid & 25k & 22  &  31  &  32 \\[0.5ex]
& 50k & 44  &  54  &  65 \\[0.5ex]
& 100k & 91  &  109  &  125 \\[0.5ex]
\end{tabular}
\caption{\textbf{Training duration for the presented neural networks with up to three interaction passes ($T=1,2,3$) in hours.}
All models were trained with stochastic gradient descent with momentum for 3.000 epochs on an NVIDIA Tesla K40 GPU.}\label{tab:traintime}
\end{table}
\clearpage

\begin{table}
\centering
\label{tab:errors}
\begin{tabular}{lrrrrrrrrrr}
Data set & \# training examples & \multicolumn{2}{c}{T=1} & \multicolumn{2}{c}{T=2} & \multicolumn{2}{c}{T=3} \\[0.5ex]
& & MAE & RMSE & MAE & RMSE & MAE & RMSE \\[0.5ex]
\hline &&&&&&& \\ [-1.5ex]
GDB-7 &  5768 & $ 1.28 \pm 0.04 $ & $ 1.99 \pm 0.14 $  & $\mathbf{ 1.04 \pm 0.02 }$ & $\mathbf{ 1.43 \pm 0.02 }$  & $\mathbf{ 1.04 \pm 0.01 }$ & $ 1.45 \pm 0.01 $ \\[0.5ex]
\hline &&&&&&& \\ [-1.5ex]      
   GDB-9 & 25k & $ 1.61 \pm 0.02 $ & $ 2.31 \pm 0.02 $ & $ 1.09 \pm 0.01 $ & $ 1.62 \pm 0.02 $ & $\mathbf{ 1.04 \pm 0.02 }$ & $\mathbf{ 1.53 \pm 0.02 }$ & \\[0.5ex]
& 50k & $ 1.49 \pm 0.02 $ & $ 2.14 \pm 0.03 $ & $ 0.96 \pm 0.01 $ & $\mathbf{ 1.37 \pm 0.03 }$ & $\mathbf{ 0.94 \pm 0.01 }$ & $\mathbf{ 1.37 \pm 0.01 }$ & \\[0.5ex]
&  100k & $ 1.54 \pm 0.03 $ & $ 2.17 \pm 0.04 $ & $ 0.93 \pm 0.02 $ & $ 1.33 \pm 0.03 $ & $\mathbf{ 0.84 \pm 0.02 }$ & $\mathbf{ 1.21 \pm 0.02 }$ & \\[0.5ex]

\hline &&&&&&& \\ [-1.5ex]
Benzene & 25k   & $ 0.07 \pm 0.00 $ & $ 0.10 \pm 0.00 $ & $ 0.05 \pm 0.00 $ & $\mathbf{ 0.06 \pm 0.00 }$ & $\mathbf{ 0.04 \pm 0.00 }$ & $\mathbf{ 0.06 \pm 0.00 }$ \\[0.5ex]
        &  50k & $ 0.06 \pm 0.00 $ & $ 0.08 \pm 0.00 $ & $\mathbf{ 0.04 \pm 0.00 }$ & $\mathbf{ 0.05 \pm 0.00 }$ & $\mathbf{ 0.04 \pm 0.00 }$ & $\mathbf{ 0.05 \pm 0.00 }$ \\[0.5ex]
        & 100k & $ 0.07 \pm 0.00 $ & $ 0.10 \pm 0.00 $ & $\mathbf{ 0.05 \pm 0.00 }$ & $\mathbf{ 0.06 \pm 0.00 }$ & $\mathbf{ 0.05 \pm 0.00 }$ & $\mathbf{ 0.06 \pm 0.00 }$   \\[0.5ex]
\hline &&&&&&& \\ [-1.5ex]
Toluene & 25k & $ 0.48 \pm 0.01 $ & $ 0.63 \pm 0.01 $ & $\mathbf{ 0.20 \pm 0.00 }$ & $\mathbf{ 0.28 \pm 0.00 }$ & $ 0.23 \pm 0.00 $ & $ 0.31 \pm 0.01 $  \\[0.5ex]
      &  50k  & $ 0.44 \pm 0.00 $ & $ 0.59 \pm 0.01 $ & $\mathbf{ 0.18 \pm 0.00 }$ & $\mathbf{ 0.24 \pm 0.00 }$ & $\mathbf{ 0.18 \pm 0.00 }$ & $\mathbf{ 0.24 \pm 0.00 }$ \\[0.5ex]
      &  100k & $ 0.42 \pm 0.01 $ & $ 0.56 \pm 0.01 $ & $\mathbf{ 0.16 \pm 0.00 }$ & $\mathbf{ 0.21 \pm 0.00 }$ & $ 0.17 \pm 0.00 $ & $ 0.22 \pm 0.00 $ \\[0.5ex]
      \hline &&&&&&& \\ [-1.5ex]
Malonaldehyde & 25k & $ 0.54 \pm 0.00 $ & $ 0.74 \pm 0.00 $ & $\mathbf{ 0.23 \pm 0.00 }$ & $ 0.34 \pm 0.00 $ & $\mathbf{ 0.23 \pm 0.00 }$ & $\mathbf{ 0.33 \pm 0.00 }$ \\[0.5ex]
             &  50k & $ 0.49 \pm 0.01 $ & $ 0.68 \pm 0.01 $ & $ 0.20 \pm 0.00 $ & $ 0.28 \pm 0.00 $ & $\mathbf{ 0.19 \pm 0.00 }$ & $\mathbf{ 0.27 \pm 0.00 }$ \\[0.5ex]
            &  100k & $ 0.51 \pm 0.01 $ & $ 0.70 \pm 0.01 $ & $ 0.18 \pm 0.00 $ & $ 0.25 \pm 0.00 $ & $\mathbf{ 0.17 \pm 0.00 }$ & $\mathbf{ 0.24 \pm 0.00 }$ \\[0.5ex]
\hline &&&&&&& \\ [-1.5ex]
Salicylic acid & 25k & $ 0.80 \pm 0.02 $ & $ 1.05 \pm 0.03 $ & $\mathbf{ 0.54 \pm 0.02 }$ & $\mathbf{ 0.72 \pm 0.03 }$  & $ 0.79 \pm 0.02 $ & $ 1.03 \pm 0.03 $ \\[0.5ex]
         &  50k & $ 0.73 \pm 0.01 $ & $ 0.94 \pm 0.01 $ & $\mathbf{ 0.41 \pm 0.00 }$ & $\mathbf{ 0.54 \pm 0.00 }$ & $ 0.50 \pm 0.01 $ & $ 0.65 \pm 0.01 $  \\[0.5ex]
        &  100k & $ 0.67 \pm 0.01 $ & $ 0.88 \pm 0.01 $ & $\mathbf{ 0.39 \pm 0.01 }$ & $\mathbf{ 0.51 \pm 0.01 }$  & $ 0.42 \pm 0.01 $ & $ 0.54 \pm 0.01 $ \\[0.5ex]
\end{tabular}
\caption{\textbf{Errors of neural networks with up to 3 interaction passes for various data sets and numbers of reference calculations used in training in kcal mol$^{-1}$.} Mean absolute errors (MAE), root mean squared errors (RMSE) as well as respective standard errors of the mean are printed. Additionally, the maximum error over all folds is given. Best results are printed in bold.}\label{tab:results}
\end{table}
\clearpage

\putbib[scibib]
		
\end{bibunit}

\end{document}